\documentclass[letterpaper, 10 pt, conference]{ieeeconf}  
\usepackage{hyperref}
\usepackage{amsmath,graphicx}
\usepackage{caption}
\usepackage{subcaption}
\usepackage{multirow}
\usepackage{adjustbox}
\usepackage[boxruled]{algorithm2e}

\IEEEoverridecommandlockouts                              

\title{\LARGE \bf
A Comparison of Representation Learning Methods for Dimensionality Reduction of fMRI Scans for Classification of ADHD
}

\author{Bhaskar Sen$^{1}$, \textit{Member, IEEE}
\thanks{$^{1}$ Independent Researcher,
        {\tt\small bsen@ieee.org}}%
}

\begin{document}

\maketitle
\thispagestyle{empty}
\pagestyle{empty}

\begin{abstract}
This paper compares three feature representation techniques used to represent resting state functional magnetic resonance (fMRI) scans. The proposed models of feature representation consider the time averaged fMRI scans as raw representation of image data. The effectiveness of the representation is evaluated by using these features for classification of Attention Deficit Hyperactivity Disorder (ADHD) patients from healthy controls. The dimensionality reduction methods used for feature representation are maximum-variance unfolding, locally linear embedding and auto-encoders. The classifiers tested for classification purpose were neural net and support vector machine. Using auto-encoders with four hidden layers along with a support vector machine classifier yielded a classification accuracy of 61.25\% along with 65.69\% sensitivity and 52.20\% specificity. 
\end{abstract}
\vspace{-2.5mm}
\section{Introduction}
\label{sec:intro}
\vspace{-1mm}
Statistical machine learning methods have recently permeated disciplines such as Psychiatry, which specializes in the diagnosis and treatment of neuropsychiatric disorders. The availability of large-scale functional Magnetic Resonance Image (fMRI) datasets have encouraged the application of advanced machine learning models to the diagnosis of neuropsychiatric disorders \cite{rish2009discriminative, senocdEMBC, sen2020sub}. fMRI scans measure brain activity by detecting fluctuations in blood-oxygen levels over time~\cite{brown2015fmri}. Hence fMRI is a four dimensional image which scans 3-D regions of interest over time.Brain activations are represented digitally as \emph{voxels}, the three-dimensional analogue of pixels.

Attention Deficit hyperactive disorder(ADHD) is a highly ubiquitous neuropsychiatric disorder affecting the lives of hundreds of millions of people. The existence of this disorder prevents a person leading a normal social life in the form of academic underachievement, inattentiveness, hyperactivities, unemployability. The causation of this disorder is attributed to structural abnormalities of specific brain regions \cite{valera2007meta} as well as atypical functional connectivity in brain. This encourages the application of functional magnetic resonance imaging for analysis of brain activities of ADHD patients and compare their difference with the fMRI scans of healthy controls. FMRI scans can capture the structural properties of the brain regions in the form of voxels of three dimensional image. Furthermore, an exciting and important extension of this analysis is to build a diagnostic model that can analyse the fMRI scan and predict with certain accuracy if the subject suffers from ADHD. In this paper, we will be building a classifier that can accurately predict ADHD with high sensitivity. This is an important incremental step for building a final classifier model that can predict ADHD both with high sensitivity and specificity.

One of the main challenges for analysis of fMRI scans is the high dimension of brain scans. A single fMRI scan may consist of hundreds of three-dimensional images over time, each of which is composed of approximately 500,000 voxels~\cite{sen2017extraction}. Therefore, extracting lower dimensional fMRI representations that retain discriminate features for diagnosis is an important step in the implementation of diagnosis algorithms.This paper investigates different representations of fMRI data, analyses three dimensionality reduction methods for feature representation and scrutinizes two classifiers for learning and classifications.

\section{Problem Statement}
\label{sec:problemstatement}
The high level objective of this paper is to be able to build a diagnostic model that can effectively classify ADHD patients from healthy controls. Ideally the model should have high sensitivity, specificity and classification accuracy. Sensitivity specifies the proportion of the ADHD patients classified by the classifier as ADHD patients. Specificity specifies the proportion of healthy subjects labeled by the classifier as true negatives and classification accuracy denotes the portion of the whole test-set correctly labeled by the classifier. Another objective is to find a good low dimensional feature representation for fMRI scans that can effectively be used for classification purposes.

\section{Related Work}
\label{sec:relatedwork}
Using the ADHD-200 competition dataset, which consists of 940 resting-state fMRI scans,\footnote{\url{http://fcon_1000.projects.nitrc.org/indi/adhd200}} Eloyan \emph{et al.} \cite{eloyan2012automated}(Winning team of ADHD-200 Competition) explored several different classifiers for ADHD diagnosis, including a support vector machine, gradient boosting, and voxel-based morphology. In addition, several feature extraction methods were investigated, including singular value decomposition and CUR matrix decomposition. The best classification accuracy was achieved by taking a weighted combination of these classifiers, which yielded $61.0\%$ accuracy on the test data.The specificity achieved is 94\% with associated sensitivity of 21\%.Also using the ADHD-200 competition dataset, Ghiassian \emph{et al.} \cite{ghiassian2016using} extracted histogram of oriented gradient features from fMRI scans, which were then input to a support vector machine. The classifier yielded an accuracy of $62.6\%$ on the test dataset. These two methods report the highest classification accuracy on the ADHD-200 competition dataset. Sen \emph{et al}~\cite{sen2018general} also used an automated learning of features from fMRI using an auto encoder and independent component analysis to classify ADHD.

Similar to previous works on fMRI for a compact representation of fMRI features, the proposed models will attempt to investigate the applicability of state-of-the art feature representation methods for fMRI and test their predictability for ADHD classification task.

\section{fMRI Data}
The dataset was taken from ADHD-200 competition.The dataset consists of functional (four Dimensional) and anatomical (three Dimensional) scans of 940 subjects. The functional scans were preprocessed following \cite{bellec2017neuro}.After preprocessing each of the subject's scans had size $79\times95\times68\times91$.The ADHD competition also had provided anatomical scans(3D) of subjects.  First,the functional scans were time averaged ($79\times95\times68$) to create 3D dimensional image for each subject.Then these scans were downsampled (one-fourth) using Gaussian Pyramid technique~\cite{anderson1985change}.The reduced dimensional image has $20\times24\times17$ voxels.The anatomical images were also downsampled to the size $20\times24\times17$ using Gaussian Pyramid. The values were sqeezed between 0-1 using a squeezing function.
Next,I also segmented regions of interests from fMRI scans for ADHD disease following \cite{bush2005functional} from \emph{cortex}, \emph{cingulate gyrus}, and \emph{thalamus} regions of the brain using Harvard-Oxford Atlas. \footnote{\url{http: //neuro.debian.net/pkgs/fsl-harvard-oxford-atlases.html}}. The voxel values for the regions are averaged and thus we get 12 time series of voxel values for each subject. I calculated the correlations between these time series for each subject and those values were converted to p-statistics. Thus we get 66 correlation values for each subject.

\section{Methods}
After downsampling the scans contain 8160 voxels which is a very large number of features to work with. Hence we reduced the number of dimensions of the input data. Three different non-linear dimensionality reduction techniques, i.e., i) Locally-Linear Embedding ii) Maximum Variance Unfolding and iii) Autoencoder were explored for feature representation. For all these dimensionality reduction techniques the input is an observation matrix $O^{t\times n}$ where $t$ is number of observation and $n$ is dimension of each observation.The output will be a reduced dimensional representation $O^{t\times m}$ where $m<n$.  
\subsubsection{Locally-linear Embedding}
One simple way for non-linear dimensionality reduction is to approximate each data point as a combination of its neighboring datapoints. This is an unsupervised way of dimensionality reduction.The number of neighbors is chosen by the user.This algorithm approximates a data point $\bf{x_{i}}$ as a composition of its neighborhood points $\bf{x_{j}}$ and corresponding weights $W_{i,j}$ ,$j\in N$ where $N$ is the neighborhood for $\bf{x_{i}}$. The optimization problem then becomes
\begin{equation}
\min_{\mathbf{W}} \sum_{i} |\mathbf{x_i}-\sum_{j}W_{i,j}\mathbf{x_{j}}|^2,
\end{equation}
Subject to $\sum_jW_{i,j}=1$ and $j\in N$ where $N$ is the neighborhood for $\bf{x_{i}}$.This optimization problem can be solved by solving a least-squares problem\cite{roweis2000nonlinear}.

\subsubsection{Maximum Variance Unfolding}
Maximum Variance Unfolding is a non-linear dimensionality reduction technique that maps the data to a non-linear feature space that is defined by a kernel function. The mapping attempts to unfold the data onto a lower dimensional manifold. We assume that the fMRI voxels lie on a lower dimensional manifold, which the algorithm will attempt to recover. In order to stretch the underlying lower dimensional manifold, the feature space should maximize the distance between two neighboring points while keeping the locality constraint intact. Formally, if $\mathbf{x}_{i}$ is a datapoint and $\mathbf{x}_{j}$ $\in N$ where $N$ defines neighborhood of $\bf{x_{i}}$ and their corresponding feature-space representation is $\Phi_{i}$ and $\Phi_{j}$ then the problem of manifold learning becomes
\begin{equation}
\max_{\mathbf{\phi}} \sum_{i,j}|\Phi(\mathbf{x}_i)-\Phi(\mathbf{x}_j)|^2,
\end{equation}
subject to the constraint that $|\Phi(\mathbf{x}_i)-\Phi(\mathbf{x}_j)|^2=|\mathbf{x}_i-\mathbf{x}_j|^2$ and $\sum_{i}\Phi(\mathbf{x}_{i}) = 0$ for all $i,j$\cite{weinberger2004learning}. This optimization problem can be efficiently solved by semidefinite programming techniques. 
\subsubsection{Autoencoder}
High-dimensional data can be converted to low-dimensional codes by training a multilayer neural network with a small central layer to reconstruct high-dimensional input vectors. Two methods for optimization, i) Conjugate Gradients(CG) and ii) Limited Memory BFGS were utilized.
Usage of CG and L-BFGS for optimising weights in a deep network was empirically studied in \cite{le2011optimization}. In this paper, we investigated the usage of these two optimization as a fine tuning step after pretraining using contrastive divergence(CD). Using intelligent guessing of weights from contrastive divergence(CD),L-BFGS and CG are expected to find the solution fast. The transfer function used in each layer was sigmoid.Layerwise the objective function is convex but as we stack the layers,the objective function becomes non-convex.Hence an intelligent guess for the initial weight becomes important. 
The autoencoder was tested for non-sparse representation as well as columnwise sparse representation.

\subsection{Feature Selection}
Feature selection becomes important if we have very low variation of a feature for all data. Before training a learner, we used PCA to choose number of required features.For doing that we used five fold cross validation to choose number of principal components based on classification accuracy on validation set.

\subsection{Classifer}
For classification, both Support Vector Machine and Neural Net classifiers were tried. But Support Vetor Machine classifier was chosen based on the reason given in ~\ref{subsec:Choice of Classifer}. For evaluating the classification models, accuracy and sensitivity were given specific attention since diagnostic process involves reducing false positives. However, accuracy, specificity, sensitivity, specificity and J-statistics were shown for final results.  

\section{Results}
\label{sec:results}

The features were extracted using a two layer autoencoder (VisibleNodes$\rightarrow$30$\rightarrow$Output nodes). The features in central layer were used for learning an SVM classifier(with rbf kernel with $\sigma=1$). The five-fold cross validation accuracy is given in Table~\ref{tab:cvresults}.
\begin{table}[h]
\caption{5-fold cross-validation results for ADHD classification (healthy, ADHD) using various dataset}
\label{tab:cvresults}
\begin{center}
\begin{sc}
\resizebox{1\columnwidth}{!}{\begin{tabular}{c c c c c}
\hline
Dataset & 5-Fold CrossValidation Accuracy \\
\hline
Time Averaged fMRI Data & \textbf{53.33}\% \\
Anatomical data  & 51.80\%  \\
fMRI Data+Correltion Matrix & 52.08\%  \\
Anatomical data+Correlation Matrix & 51.93\%  \\
\hline
\end{tabular}}
\end{sc}
\end{center}
\end{table}
As clearly seen from the accuracy, time averaged fMRI data outperforms others for accuracy.

Number of layers in the autoencoder was chosen using layerwise pretraining and backpropagation (Conjugate gradient with three line searches). Number of iteration in pretraining was fixed at 10 and number of iteration during back-propagation was fixed at 5. These numbers (iterations) were empirically chosen that gave best accuracies on 5-fold cross validation. The features (70 for all the experiments) were used to train a SVM classifier with rbf kernel. The sigma values were tested from 0.01 to 10.$\sigma=0.6$ gave the best 5-fold cross validation accuracy and hence the result is shown just for $\sigma=0.6$. Here hidden layers upto central layer has been shown. The other half reciprocates the first half. 

\begin{table}[h]
\caption{Accuracy and Sensitivity Using Different Layers.}
\label{tab:hidlayers}
\begin{center}
\begin{sc}
\begin{tabular}{c c c}
\hline
Hidden Nodes & Accuracy &Sensitivity\\
\hline
8160$\rightarrow$4000$\rightarrow$1000$\rightarrow$500$\rightarrow$70 & \bf{58.13}\% &50\%\\
8160$\rightarrow$1000$\rightarrow$500$\rightarrow$70  & 55.60\%  &\bf{65}\%\\
8160$\rightarrow$200$\rightarrow$70 & \bf{57.01}\%  &\bf{62.26}\%\\
\hline
\end{tabular}
\end{sc}
\end{center}
\end{table}
As stated in~\cite{sen2018general}, the distribution of the ADHD and Healthy patients in the dataset is skewed. Hence 8160$\rightarrow$200$\rightarrow$70$ \rightarrow$200$\rightarrow$8160 met the criteria even though the autoencoder with nine hidden layer gave best accuracy but poor sensitivity. On the other hand the autoencoder with seven hidden layers gave very high sensitivity but poor accuracy. The sensitivity of the representation decreases for a very deep layer.The accuracy is high because it just labels most as majority class. This answer is not definite though as we see some fluctuations in sensitivity with increase of depth of layers.Ideally many other possible depths should have been tested. Also experiments can be done with the width of each hidden layer to find a more definite answer. We used nodes of central hidden layer of $8160\rightarrow 200 \rightarrow 70\rightarrow 200\rightarrow 8160$ as features of the SVM classifier. Effectively, 70 features were used for testing classifier accuracy along with sensitivity.Here also for learning and classification an SVM classifier along with rbf kernel was chosen.For testing purpose with Conjugate Gradient and L-BFGS,$\sigma=0.6$ was chosen.The result is given in~\ref{tab:opti}
\begin{table}[h]
\caption{Accuracy and Sensitivity Using Different Optimization}
\label{tab:opti}
\begin{center}
\begin{sc}
\begin{tabular}{c c c}
\hline
Hidden Nodes & Accuracy &Sensitivity\\
\hline
CG(With 3 Line Searches) & 57.01\% &62.26\%\\
L-BFGS& \bf{57.90}\%  &\bf{63.01}\%\\
\hline
\end{tabular}
\end{sc}
\end{center}
\end{table}
Here clearly L-BFGS gave better accuracy as well as sensitivity.Hence the hypothesis that L-BFGS should give better representation was true.

Next, testing different sparse representations was necessary as the features learned with no regularization had some features always activated as shown. This was done introducing two sparse regularizers in the hidden layer.If $X$ is our data matrix and $\phi$=$f(XV+\bf{1.a^{T}})$ is the hidden representation of the data then,

\begin{equation}
\label{eqn:reg1}
{\tt Regularizer1} =\arrowvert\arrowvert \phi \arrowvert\arrowvert_{1,1}=1^{T}\phi1.
\end{equation}
and 
\begin{equation}
\label{eqn:reg2}
{\tt Regularizer2}=\frac{1}{2}\arrowvert\arrowvert (\frac{1}{t}{\bf 1}^{T}\phi)-\rho {\bf 1}^{T} \arrowvert\arrowvert_{2}^{2}.
\end{equation}
where t is number of observations and $\rho=0.1$. Regularizer1 in Eq.~\ref{eqn:reg1} introduces sparsity throughout the representation matrix.Whereas Regularizer2 in Eq~\ref{eqn:reg2} induces columnwise sparsity in the hidden representation matrix.Hence Regularizer2 ~\ref{eqn:reg2} encourages distinctive feature for different subjects.The fine tuning using L-BFGS was done layerwise and stacked to form a two layer model.
Their result on classification accuracy was also compared with feature representation without any sparsity inducing regularizer.The results of the classification accuracy for no regularizer,regularizer1 regularizer2 is shown in Table~\ref{tab:opti1}
\begin{table}[h]
\caption{Accuracy Using different Sparse Representations}
\label{tab:opti1}
\begin{center}
\begin{sc}
\begin{tabular}{c c}
\hline
Regularizers & Accuracy\\
\hline
Noregularization& \textbf{57.90}\% \\
Regularizer1& 56.25\%  \\
Regularizer2 &\textbf{58.50}\%\\
\hline
\end{tabular}
\end{sc}
\end{center}
\end{table}

\begin{figure}[h]
	\centering
	\includegraphics[width=0.5\textwidth]{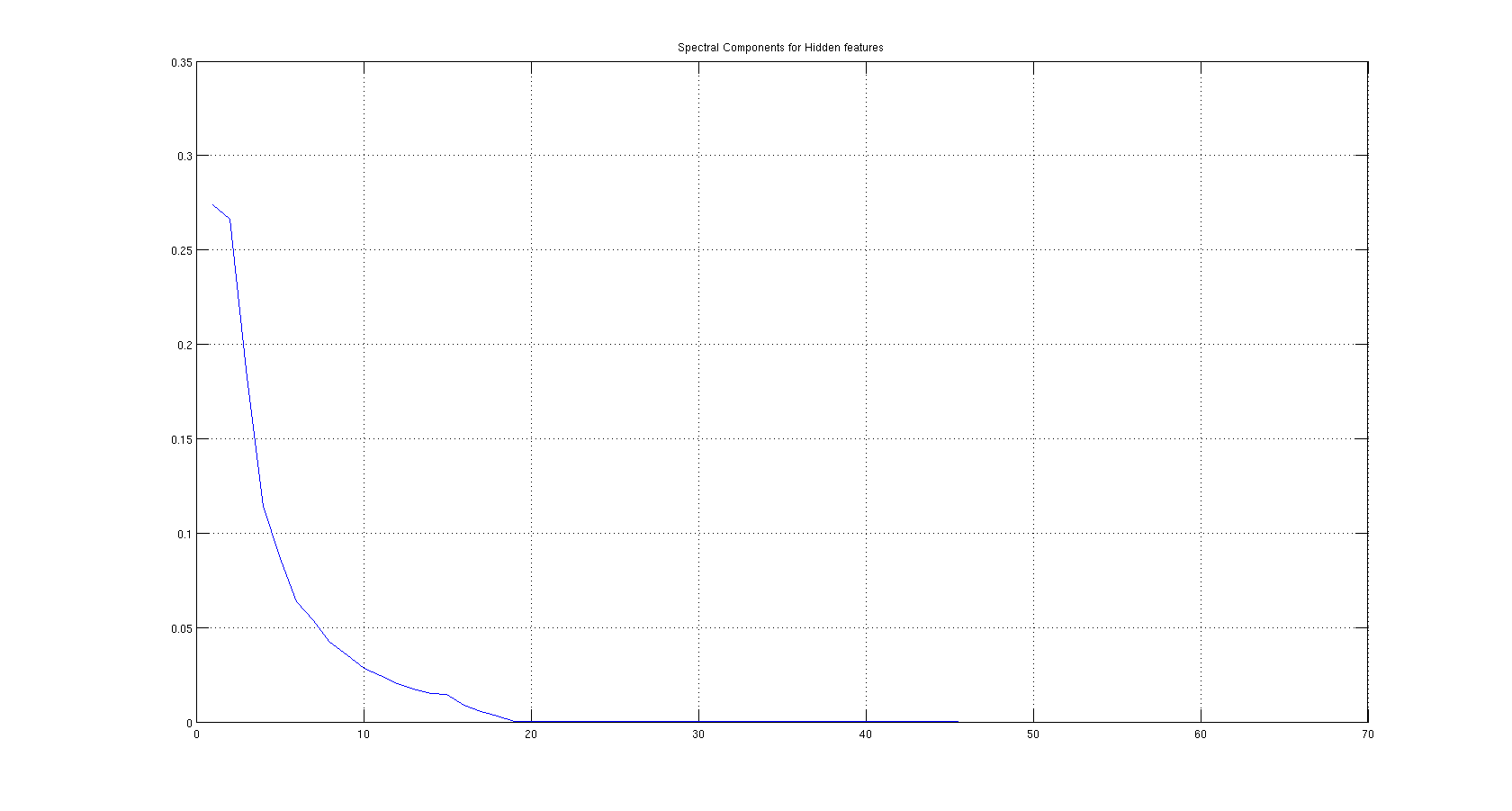}
    \caption{Spectral Values for Hidden Layer Features}
    \label{fig:al2}
\end{figure}

From the result,the columnwise sparsity inducing matrix had the best accuracy for classification.This clearly make sense because if we compare the features and basis images learned from No-regularization and Regularizer2 ~\ref{fig:al2},the columnwise sparsity inducing matrix gave a better representation of features. On the other hand Regularizer1 fails to achieve good accuracy.

Next, we tested autoencoder representation with representation got from Locally Linear Embedding(LLE) and Maximum-Variance Unfolding(MVU) method.LLE was implemented following  \footnote{\url{http://http://www.cs.nyu.edu/~roweis/lle/code.html}}.  Also MVU was implemented using  \footnote{\url{http://http://www.cse.wustl.edu/~kilian/code/files/mvu2012.zip}}.For comparison purpose, the reduced dimension was  fixed at 70. For LLE and MVU,number of neighbors chosen empirically was $N=30$. For classifier an SVM classifier with rbf kernel($\sigma=0.6$) was used. The five fold cross validation accuracy is given below.
\begin{table}[h]
\caption{Accuracy and Sensitivity Using different Representation Methods}
\label{tab:opti2}
\begin{center}
\begin{sc}
\resizebox{1\columnwidth}{!}{\begin{tabular}{c c c c c}
\hline
Hidden Nodes & Accuracy &Sensitivity &Specificity &J-Stat\\
\hline
LLE & 55.65\% &43.38\% &\textbf{65.66}\% &8\%\\
MVU& 52.21\%  &57.55\% &64.93\% &\textbf{22\%}\\
Autoencoder &\textbf{58.50}\% &\textbf{63.21}\% &43.28\% &6\%\\
\hline
\end{tabular}}
\end{sc}
\end{center}
\end{table}

The autoencoder outperforms the other two methods.It achieves high accuracy and sensitivity but low specificity.On the other hand LLE gives better representation than MVU even though sensitivity is better for MVU.

The 70 features,which represents the input data,contain some features which do not vary for different subject.This is clear from the features and the basis image received from the autoencoder with nodes 8160$\rightarrow$200$\rightarrow$70(For L-BFGS). We chose to use PCA to select important features among these. From the spectral components of these features, as shown in Fig.~\ref{fig:al2},

The most useful information (above 90\%) among the features lies in first few components.Hence we used PCA to find those components and used five-fold cross validation for choosing best number of components to be used for my algorithm.For training classifier again SVM classifer was used with rbf kernel.Here  five-fold cross validation was used to choose best $\sigma$.Result is shown for $\sigma=0.1$

\begin{table}[h]
\caption{Using PCA Components}
\label{tab:PCA}
\begin{center}
\begin{sc}
\resizebox{1\columnwidth}{!}{\begin{tabular}{c c c c c}
\hline
PCA Components & Accuracy &Sensitivity &Specificity &J-Stat\\
\hline
2 & 58.33\% &\textbf{84.29}\% &21.12\% &5\%\\
5& \textbf{61.25}\%  &65.69\% &52.20\% &\textbf{17\%}\\
7 &57.31\% &59.75\% &\textbf{54.21}\% &13\%\\
10 &60.01\% &63.37\% &48.51\% &11\%\\
15 &58.33\% &54.21\% &53.55\% &7\%\\
\hline
\end{tabular}}
\end{sc}
\end{center}
\end{table}
Here choosing first five components gave best accuracy result. As shown in ~\ref{tab:PCA1} the final algorithm achieved five-fold cross validation accuracy of 61.25\%  which is better than \cite{eloyan2012automated} but slightly less than \cite{ghiassian2016using}.The algorithm achieved Sensitivity and J-Stat better than \cite{eloyan2012automated} but Specificity was lower.
\begin{table}[h]
\caption{Comparison of the proposed Algorithm to ADHD-200 Competition}
\label{tab:PCA1}
\begin{center}
\begin{sc}
\resizebox{1\columnwidth}{!}{\begin{tabular}{c c c c c}
\hline
Algorithm & Accuracy &Sensitivity &Specificity &J-Stat\\
\hline
Our Result & \textbf{61.25}\% &\textbf{65.69}\% &52.20\% &\textbf{17\%}\\
ADHD-200 Result& 61.0\%  &21\% &\textbf{94}\% &15\%\\
\hline
\end{tabular}}
\end{sc}
\end{center}
\end{table}

Even though my model achieved close to state of the art performance for diagnosis of ADHD,it is far from satisfactory to be used for any practical purpose.The reason behind the lacklustre performance might be lack of important information in the fMRI scans for detecting ADHD.It might be possible that we are losing effective information while downsampling the fMRI scans.Hence, using the full images as the input of autoencoder can be tested.But this requires very high computation power.

\subsection{Choice of Classifer}
\label{subsec:Choice of Classifer}
For choosing the classifer to be used,initially we had decided to test two classifiers:Neural Net and Support Vector Machine.But initial tests with Neural Net classifier showed that it was just following the majority class classification.Also using a neural net classifer did not permit to visualise the features learned.Hence we decided to use autoencoder for learning features and comparing  with other feature representation method and use SVM for classification.

\section{Conclusion}
\label{sec:conclusion}
The development of automatic ADHD diagnostic algorithms from fMRI data is a challenging task. The application of statistical pattern recognition algorithms to this problem currently yield insubstantial results, rendering these classification systems unfit for practice in the health-care industry. However, much research is being done to improve these results and search for discriminating features for classifying ADHD amongst the plethora of voxel values present in a single fMRI scan. Apart from systems that aggregate fMR images over time to produce a single three-dimensional image of the brain that is then used for classification, in this paper we explored the applicability of different feature representation techniques that can be used for effective representation of fMRI scans. Specifically, we used ADHD-200 competition dataset for learning features using Autoencoder, Locally Linear Embeddings and Maximum Variance Unfolding.My results indicate that autoencoders followed by principal component analysis and support vector machine classifer, yields the best results with 61.25\% 5-fold cross-validation accuracy.

Still, there is much work to be done in this area. Recurrent neural networks or convolutional neural network, should be explored in future work. Moreover, we propose that locating lower dimensional sets of fMRI features that retain discriminating power for ADHD classification is the heart of the problem and that the majority of future work should focus on this task.

\vspace{-2mm}
\bibliographystyle{IEEEbib}
\bibliography{strings,refs}

\end{document}